\documentclass[11pt]{article}
\usepackage{fullpage}
\usepackage[table,xcdraw]{xcolor}

\renewcommand{\thesection}{\arabic{section}}
\renewcommand{\thesubsection}{\thesection.\arabic{subsection}}
\renewcommand{\thesubsubsection}{\thesubsection.\arabic{subsubsection}}

\usepackage{titlesec}
\titleformat{\section}[block]{\Large\bfseries}{\thesection}{1em}{}
\titleformat{\subsection}[block]{\large\bfseries}{\thesubsection}{1em}{}
\titleformat{\subsubsection}[block]{\normalsize\bfseries}{\thesubsubsection}{1em}{}

\usepackage{graphicx} 
\usepackage{natbib}
\usepackage{hyperref}
\usepackage{color}

\usepackage{graphicx, centernot, amsmath, bbm, booktabs, multirow, array, capt-of, varwidth, amstext, tabularx}
\usepackage{enumitem}
\usepackage{subcaption}
\usepackage{cleveref}
\usepackage[utf8]{inputenc}
\usepackage{dirtytalk}
\usepackage{csquotes}
\usepackage{graphicx}
\usepackage{amsmath,amsthm,amsfonts}
\usepackage{lipsum}
\usepackage[utf8]{inputenc}
\usepackage{hyperref}
\usepackage{float}
\hypersetup{
    colorlinks=true,
    linkcolor=blue,
    filecolor=magenta,      
    urlcolor=cyan,
    pdftitle={Overleaf Example},
    pdfpagemode=FullScreen,
    }
\usepackage{tikz}
\usetikzlibrary{bayesnet}

\usepackage[many]{tcolorbox}   
\definecolor{main}{HTML}{5989cf}    
\definecolor{sub}{HTML}{cde4ff}     

\tcbset{
    sharp corners,
    colback = white,
    before skip = 0.2cm,    
    after skip = 0.5cm      
}                           

\newtcolorbox{boxA}{
    fontupper = \bf,
    boxrule = 1.5pt,
    colframe = black 
}

\newtcolorbox{boxB}{
    fontupper = \bf\color{main}, 
    boxrule = 1.5pt,
    colframe = main,
    rounded corners,
    arc = 5pt   
}

\newtcolorbox{boxC}{
    colback = sub, 
    boxrule = 0pt  
}

\newtcolorbox{boxD}{
    colback = sub, 
    colframe = main, 
    boxrule = 0pt, 
    toprule = 3pt, 
    bottomrule = 3pt 
}

\usepackage{tabularx}
\usepackage{booktabs}   
\usepackage{colortbl}   
\usepackage{multirow}   

\title{Bridging Predictions and Interventions: An Integrated Framework for Automated Decision-Systems}
\author{
Inioluwa Deborah Raji\footnotemark[1],~Lydia T.~Liu\thanks{These authors contributed equally to this work.}~, Angela Zhou\footnotemark[1]~,\\ Luke Guerdan,  Jessica Hullman, Daniel Malinsky, Bryan Wilder, Simone Zhang, \\
Hammaad Adam,  Amanda Coston, Ben Laufer,  Ezinne Nwankwo,  Michael Zanger-Tishler,\\ 
Eli Ben-Michael, Avi Feller, Talia Gillis, Shion Guha, Daniel Ho, Lily Hu,
 Kosuke Imai,\\ Sayash Kapoor, Joshua Loftus, Razieh Nabi,  Juan Carlos Perdomo, Matthew Salganik,\\ Mark Sendak, 
Berk Ustun,  Suresh Venkatasubramanian, Angelina Wang, Ashia Wilson 
}

\titleclass{\subsubsection}{straight}
\titleformat{\subsubsection}%
  [hang]
  {\normalfont\large\itshape}
  {\thesubsubsection}
  {1em}
  {}
  []
  
\date{\today}

\usepackage{endnotes}

\usepackage{amsmath, amsthm, amssymb, amsfonts}
\usepackage{algpseudocode}
\usepackage{algorithm}

\algtext*{EndIf} 
\algtext*{EndWhile} 
\algtext*{EndFor} 
\algtext*{EndFunction} 
\algtext*{EndProcedure} 

\usepackage{graphicx}
\usepackage{bm}
\newcommand\E{\mathbb{E}}

\usepackage{wrapfig}

\usepackage{hyperref}

\usetikzlibrary{shapes,decorations,arrows,calc,arrows.meta,fit,positioning}
\tikzset{
    -Latex,auto,node distance =1 cm and 1 cm,semithick,
    state/.style ={ellipse, draw, minimum width = 0.7 cm},
    point/.style = {circle, draw, inner sep=0.04cm,fill,node contents={}},
    bidirected/.style={Latex-Latex,dashed},
    el/.style = {inner sep=2pt, align=left, sloped}
}

\begin{document} 


\maketitle

\begin{abstract}


Automated decision systems (ADS) leverage predictions about individual future outcomes to inform consequential decision-making in organizational settings. 
Across various settings --- including criminal pretrial release, clinical triage, student support, and more --- it is often assumed that improved predictive accuracy is the priority consideration in determining better downstream outcomes upon the deployment of ADS. 
In practice, real-world case studies reveal that this is far from the case: introducing individual predictions into decision-making modifies organizational workflows, assessment, and decision-making processes in ways that require a complete re-consideration of our approach to the design, evaluation, and deployment of ADS. 
As a result, this Perspective develops an integrated framework for studying ADS in social systems, shifting current priorities from a purely prediction-based paradigm towards an intervention-oriented view 
that accounts for real-world conditions. 
Our aim is to improve our understanding of ADS and more meaningfully anticipate its downstream societal and organizational consequences. 


\end{abstract}

\section{Introduction}

Automated decision systems (ADS) leverage statistical pattern recognition in historical data to make predictions about individuals. Developed using computational methods ranging from simple regression to more complex generative modeling, the goal of ADS is to inform bureaucratic decision-making~\citep{richardson2021defining}.
Organizations cite common reasons for their adoption of ADS: standardized predictive assessments can help structure complex casework, reduce arbitrary variability in decisions, and achieve strong predictive performance in difficult tasks \citep{mdrcEvaluationPretrial,njcourts}.

Yet a central paradox persists: despite the increasing ubiquity of ADS in public life, evidence that its deployments improve social outcomes is slim or inconsistent at best, revealing a substantial gap between predictive performance and real-world impact.
While ADS developers report 65--70\% accuracy rates in recidivism prediction on undisclosed evaluation data \citep{larson2016compas},
it remains unclear whether RATs reduce  detainment rates while ensuring public safety as claimed\citep{angwin2016machine,green2020false}. 
Similarly, the widely deployed Epic Sepsis Model (ESM) reported a high area-under-the-curve (AUC) performance of over 0.80; yet in external validation, the clinical decision support tool failed to identify 67\% of thousands of hospitalized patients with life-threatening sepsis, while simultaneously flagging too many patients without sepsis --- prompting user distrust and alert fatigue ~\citep{wong2021external}. Predictive tools used in early-warning educational systems to identify students at risk of dropout have had equally mixed results \citep{feathers2021racepredictor,feathers2023takeaways}, with some sites documenting significant improvements in graduation rates \citep{rossmanMAAPSAdvisingExperiment2023}, while other districts finding weak to null effects on student success \citep{perdomo2023difficult}. This landscape of mixed results highlights persistent gaps in our understanding of how ADS manifest in practice. 


Across domains, myopically focusing on standalone ADS \emph{model} properties, such as predictive accuracy \citep{kleinberg16guide,kleinberg2015prediction}, rarely translates straightforwardly into any meaningful judgment of improvement to \emph{downstream} outcomes. 
Upon deployment, complexities abound: ADS are not just stand-alone artifacts, but operate as bureaucratic interventions, changing how organizations make decisions. 
In this Perspective, 
we bridge analysis of ADS ~\emph{prediction} accuracy with an expanded understanding of how ADS act as bureaucratic ~\emph{interventions}. 
More specifically, when organizations deploy ADS, they also enact an organizational \textit{policy change} by modifying processes of individual \textit{assessment} and \textit{decisions}. 
Often focused on limits to predictive accuracy, those that design and deploy ADS typically overlook the importance of how organizations operationalize these tools, i.e., how they adopt, display, and act on predictions. Expanding on community-level findings~\citep{liu2025bridging}, our integrated view can inform new methods of ADS design, evaluation, and implementation to bridge the crucial gap between our understanding of model properties and our judgment of the associated real world consequences.

\section{Problem setup}

ADS are conceptualized to improve or standardize population-level outcomes (e.g. student dropout, patient readmission, recidivism rate) by informing an organization's decision-makers (e.g. academic advisors, doctors and nurses, or judges) of the underlying risk of an individual decision subject \citep{kleinberg2015prediction}. 
This population of \emph{individual decision subjects} (e.g. students, patients, pre-trial defendants) is often represented by covariate information $X_i$ (e.g. demographics and history). Although each subject eventually experiences a measured outcome, $Y_i$ (e.g. drop out, re-admission, recidivism), this outcome is often not observable at the point of decision-making. Therefore, the most common ADS is a predictive model $R(X)=\E[Y\mid X]$ that predicts the outcome $Y_i$ for each individual decision subject at the point that a decision is being made about that individual. 

\begin{figure}[ht!bp]
\centering\includegraphics[width=0.8\linewidth]{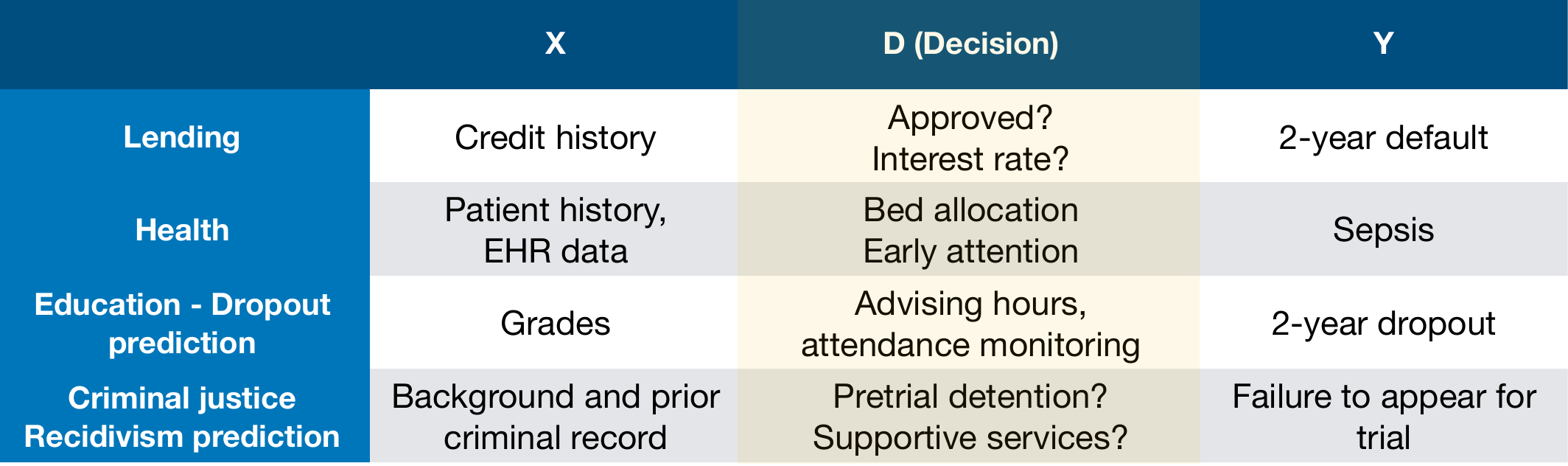}
    \caption{Example of predictive ADS and the decisions they inform.}
    \label{fig:examples}
\end{figure}

Typical conceptions of ADS focus on $(X,R,Y)$ alone. 
Yet, as previously discussed, predictive information rarely impacts actual individual outcomes in isolation. Rather, predictions inform downstream decision-makers that in turn affect outcomes. \Cref{fig:examples} illustrates how this arises across different domains: dropout-prediction directs the decision of who receives additional student advising resources and that in turn is what impacts actual student dropout rates -- in the same way, diagnostic prediction informs decisions to triage and treat; and predicted recidivism risk can inform decisions for detention, carceral resource allocation \citep{barabas2018interventions}, supervisory conditions \citep{schuman2019supervised}, or court diversion programs. 

Introducing predictions into operational workflows changes consequential decision-making processes. Therefore, in our integrated view, we introduce the additional considerations of ~\emph{assessment} --- i.e. how predictive information shapes the sorting, categorical arrangement or ranking of individual subjects --- and \emph{decisions} --- i.e. intervention actions available to decision-makers that can have a direct material influence on individual decision-subject outcomes. 



\paragraph{Assessment, $\hat Y$.} 
``Assessment'' maps observed individual-level information (i.e. covariates, $X_i$), to simpler categories that can be used to inform future decisions. 
In the sepsis prediction example, the assessment is the binary ``high-risk label'' that pages a nurse or not. In drop-out prediction, the assessment is labeling a student ``at-risk of dropout''. 

Designing a meaningful assessment involves designing the categories and the boundaries between them. Although on-the-ground decision-makers already perform discretionary categorization when taking into account input signals as well as individual beliefs and organizational guidance, ``assessment'' can also encompass algorithmic categorization. 
Common practices of algorithmic categorization include thresholding predictive risk scores to high- or low-risk labels $\hat Y = \mathbb{I}[R(X)>t],$ that then drive the form of decisions for all within a category. 

\paragraph{Decision, $D$.} 
Decisions involve making a choice among the available actions that can be taken by an individual decision maker, typically affecting individual decision subjects. 
In the education setting, schools guide guidance counselors on which students receive additional resources such as  1:1 student counseling. In sepsis management, hospitals select patients for early intervention. In pre-trial adjudication, judges decide whether or not to release or detain individuals. 

Predictive output $\hat{Y}$ may inform decision-makers without directly determining which actions should be taken in order to optimize decision utility or downstream outcomes~\citep{liu2024actionability}. When decisions have an unknown causal effect on outcomes, we can use the potential outcomes framework to describe an individual's outcomes under different decisions \citep{rubin2005causal}. 
Each individual is described by their individual attributes (i.e. covariates $X$), realized decision(s) $D$, and realized outcome(s) $Y$. 
Each individual has a random vector of potential outcomes $Y(d)$, $d \in \mathcal{D}$, and $Y(d)$ indicates the outcome realized under decision $D=d$. In practice, we only observe the realized outcome $Y = Y(D)$, and must estimate the other potential outcomes under different decisions than the decision that was actually made for this individual --- this gives rise to the
``fundamental problem of causal inference." For binary decisions, the average treatment effect (ATE) describes the population-level average effect of assigning a particular decision or not to all decision subjects: 
$$ \textrm{ATE} := \E[Y(1)-Y(0)]. $$

When decisions have different effects for different subjects, we consider the treatment effect to be heterogeneous. We can thus define a conditional average treatment effect (CATE) to describe the effect of a decision on someone with specific covariates $X$:

$$
\textrm{CATE}(X) := \E[Y(1)-Y(0) \mid X]. 
$$

\paragraph{Policy change, $Z$.}
ADS are not deployed in a vacuum: institutions introduce ADS into an organizational process, enacting a ``policy change'' from status quo practice, to an updated workflow of ADS-mediated decision-making. 
Organizations set the criteria for ADS deployment, determining not just whether to deploy ADS or not, but also which cases are impacted and how. Organizations also define the structure of decisions, $D,$ by determining what actions are available to be taken by the decision-maker, as well as providing the guidance that informs the choice between actions. 
For example, a policy change in criminal justice might be whether or not a judge sees a particular PSA risk score (e.g. Kansas law HB463 requires judges to heed risk score recommendations or otherwise file a recorded reason for making an alternative decision), setting the risk levels (e.g. multi-level high/low/medium, vs binary low/high) and the available set of actions a judge can take in response (e.g. detain vs release vs supervise). In healthcare, the policy change might involve a medical association or hospital's guidance on how to respond to high predicted sepsis risk, including whether or not a risk score is sufficient evidence to take diagnostic or triage action.


We use $Z$ to denote shifts in the policy regime organizations set to impact individual-level decisions. 
When ADS are introduced into an organizational setting at some fixed time, $Z\in\{0,1\}$ can denote the time before or after ADS deployment. 
As organizations can also determine which decisions are subject to an ADS recommendation, $Z_i \in \{0,1\}$ may also indicate whether or not an individual decision-maker sees a prediction score for a particular individual decision subject. Policy change enables estimating causal effects of ADS on decisions by comparing decisions made with or without ADS.  





\paragraph{An integrated perspective.}

\begin{figure}
    \centering
    \includegraphics[width=0.6\linewidth]{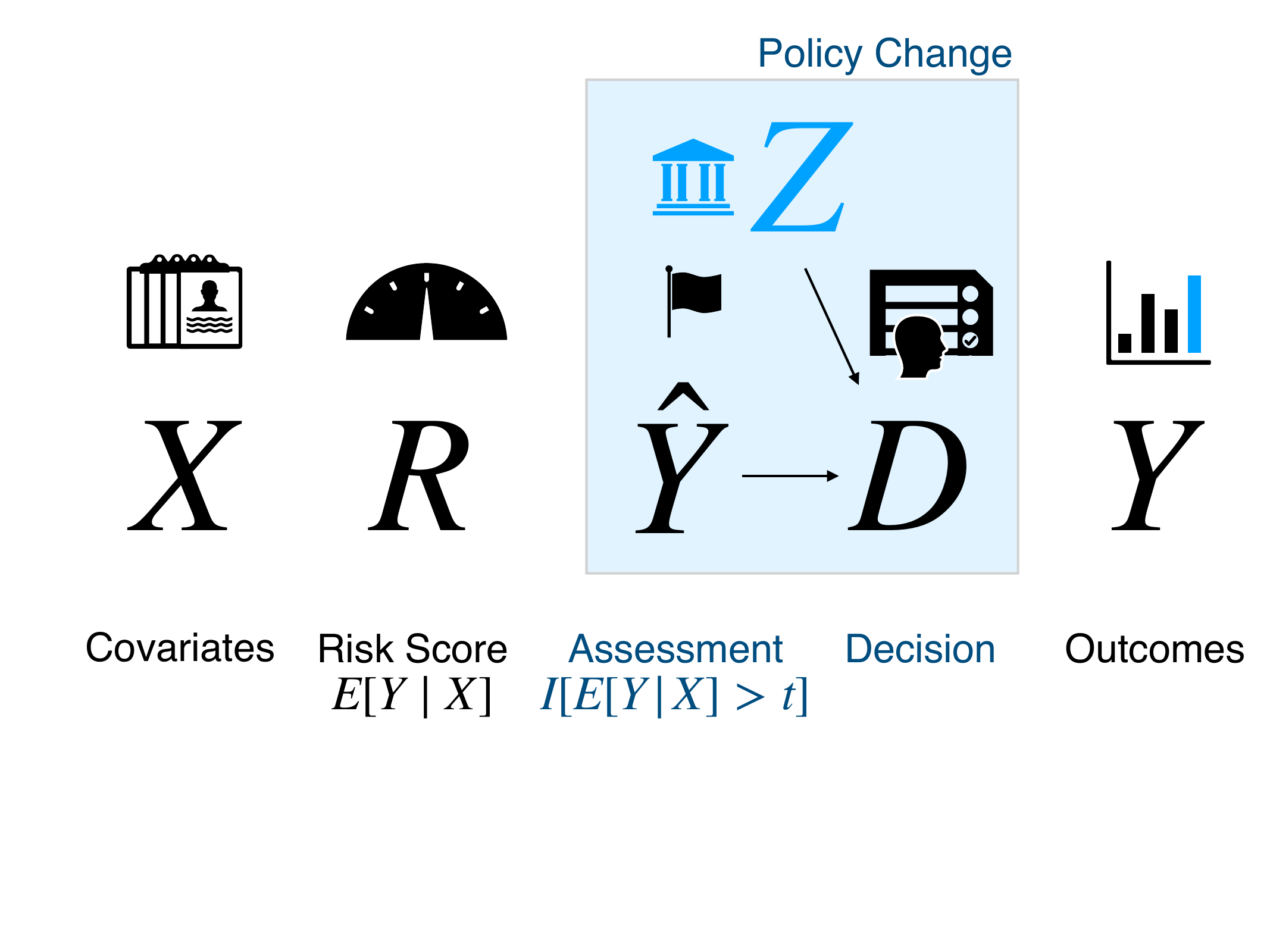}
    \caption{
    \textbf{From prediction to intervention in automated decision systems.} Predictive risk scores \(R\), generated from individual data covariates \(X\), does not affect outcomes \(Y\) directly. It operates through assessments \(\hat{Y}\), which leads to decisions \(D\) under an institutional policy change \(Z\), such as the introduction of an ADS into an existing bureaucratic workflow.}
    \label{fig:conceptual-diagram}
\end{figure}

Currently, the study of the ADS pipeline --- from predictions, to judgment, decisions and outcomes --- is distributed across different disciplinary communities (See \Cref{fig:conceptual-diagram}). As a result, individual studies typically consider \textit{distinct subsets} of our expanded viewpoint, and overlook specific factors worth bringing to the forefront in implementation settings. 
For example, even for the same ADS -- pretrial Risk Assessment Tools (RATs) -- multiple studies from different perspectives might illuminate distinct angles to the broader question: \textit{do algorithmic pretrial risk assessments improve efficiency and/or equity in pretrial judgments?} 

\begin{itemize}
    \item $(X, R, Y)$: Benchmarking studies of COMPAS focus on the predictive error and classification metrics associated with RATs~\citep{angwin2016machine, bao2021s, dieterich16compas}, such as prediction rates, true positive rates, and other classification metrics. 
    \item $(R,\hat{Y},D)$: Behavioral lab studies of responses to RATs analyze how individuals make decisions given risk score information, i.e. detention decisions $\mathbb{P}(D=1 \mid R(X))$ given a risk score,  assessment categories, $\mathbb{P}(D=1 \mid \hat Y=1)$ ~\citep{lai2021towards}, or under different choices in presentation and definition of categories~\citep{yin2019understanding}. 
    \item $(Z,D)$: Observational \cite{albright2019if,angelova2025algorithmic} and behavioral lab studies~\cite{green2019disparate} estimate racial disparities in decision outcomes $D(z)$ made with or without a risk score ($Z\in \{0,1\}$) to investigate the role of \emph{human discretion} in overriding algorithmic recommendations to determine final decisions. 
    \item  $(Z,D,Y)$: Experimental evaluation studies such as \cite{ben2025does,imai2023experimental} study the impact of showing the risk score on decisions that do impact downstream outcomes. 
    
    
    \item $(D,Y)$: Substantive studies in the social sciences evaluate the \textit{causal} impact of pretrial detention on downstream cases and individual outcomes \citep{koppel2024examining,st2024pretrial,dobbie2018effects,heaton2017downstream,leslie2017unintended,holsinger2018analyzing}, without controlling the policy change ($Z$) directly. 
\end{itemize} 

To be sure, a rich collection of prior frameworks emphasize aspects of our integrated view as well. The literature on algorithmic fairness studied how ADS optimized for population-level loss can introduce individual-level harms, especially for marginalized groups \citep{barocas-hardt-narayanan}. Furthermore, \citet{chohlas2024learning,hu2020fair,liu18delayed} and others emphasize consequentialist and social-welfare frameworks for evaluating machine learning in consequential social settings. However, as mentioned previously, the consequences of ADS deployments cannot be evaluated before the fact by scrutinizing formal properties of ADS alone -- a key contribution of this integrated view is thus an expanded technical vocabulary for analyzing deployed ADS within its pragmatic context in organizational decision-making. 

\section{Implications of an integrated view} 

We conceptualize the lifecycle of an automated decision system (ADS) in three interrelated stages: model design, evaluation science, and implementation science.  
In this section, we discuss how
we can adopt tools from human-computer interaction studies, program evaluation and organizational behavior the model design, evaluation and implementation science of ADS. 

\begin{figure}
    \centering
    \includegraphics[width=0.5\linewidth]{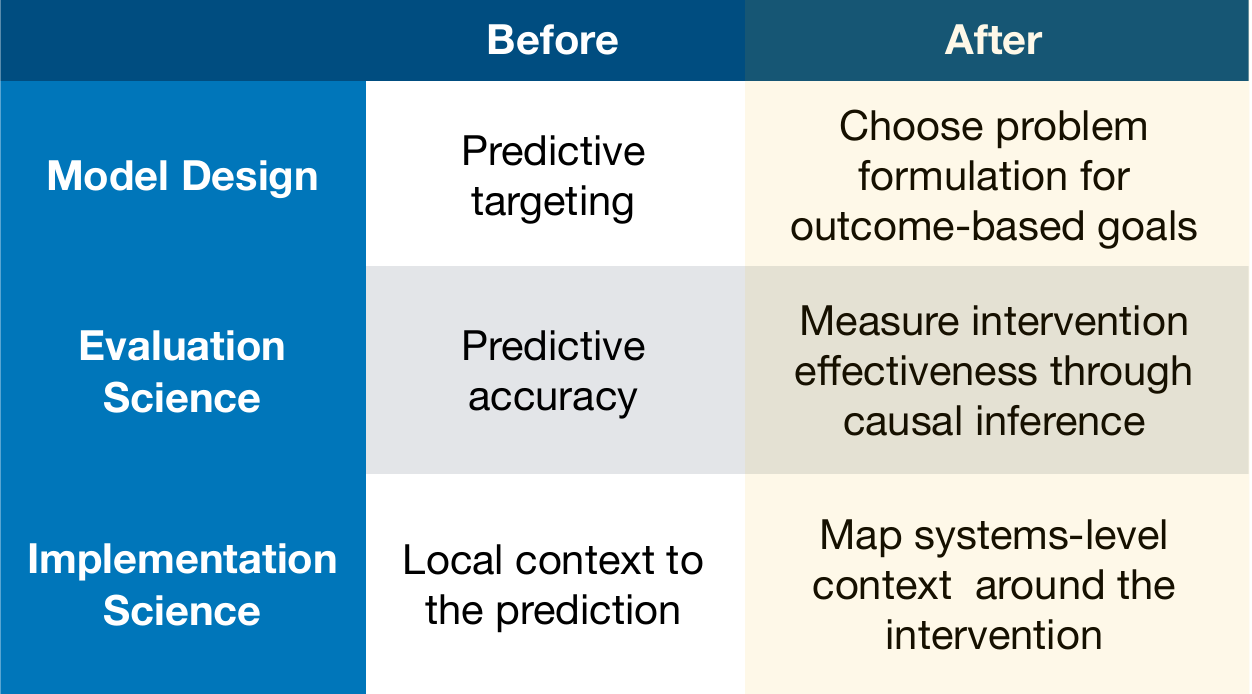}
    \caption{ADS lifecycle before and after our integrated view.}
    \label{fig:placeholder}
\end{figure}

\subsection{Model design}

\emph{Model design} refers to the problem formulation phase of developing an actual ADS tool --- that is, how to set model objectives, and algorithmically optimize those objectives. We argue that the design of ADS tools should explicitly consider alternatives to prediction modeling based on the nature of an organization's underlying decision problem.

In many cases, pure prediction can be a sufficient goal to inform meaningful and effective decision-making. However, there are just as many cases where prediction is not enough, and an interventional framing better informs model design. This all depends on the domain context of how predictions translate into decision-utility. Organizations have a loss function depending on observed decisions and outcomes $\ell(D,Y).$ When decision-makers have domain knowledge on how exactly changes in decision affect changes in outcomes, counterfactual loss can be evaluated from historical data alone, and simple prediction is enough. As a simple example, consider whether or not to bring an umbrella: since it's known that umbrellas keep you dry for sure, it's enough to predict whether or not it rains \citep{klmo15}. But when decisions have a priori unknown causal effects on outcomes,
($D \rightarrow Y(D) \rightarrow \ell(D,Y))$, optimizing
policies requires causal estimation. \citet{liu2024actionability} also studies when outcome predictions are actionable vs. not.

Organizations should compare predictive targeting with interventional improvement-based prioritization. 
The ``prediction view'' and ``policy intervention view'' triage on different objects: baseline risk under the standard prediction view, vs. improvement from decision under the intervention view. 
\begin{align*}
\pi_{\text{risk}}(X)&=\mathbb{I}[\E[Y(0)|X]>t]. \tag{``Prediction view'', targeting on baseline risk}\\
\pi_{\text{ITR}}(X) &= \mathbb{I}[ \E[Y(1)-Y(0)|X] >t] \label{eq:itr}\tag{``Policy intervention view'', optimal treatment policy}
\end{align*}
Standard triage and ranking uses predictive tools that are therefore prioritizing \textit{baseline risk}. 
In homelessness services, the VI-SPDAT assesses ``vulnerability'', i.e. the probability of returning to homelessness within two years, if not provided with additional housing support $P(Y(0) \mid X)$, and jurisdictions widely adopted it after a federal mandate to implement \textit{some} formal prioritization scheme, though it was never initially resourced to be widely validated and deployed. 
Prioritizing based on baseline risk, need or vulnerability reflects important societal values of protecting the worst-off, and it is also easier to estimate baseline rather than counterfactual risk. But when resources are limited, targeting interventions (such as who to provide with more or less intensive housing supports) to those who benefit the most can achieve the most average improvement under limited sources \citep{kube-das-19,rahmattalabi2022learning,azizi2018designing}.
Targeting based on baseline (prognostic) risk vs. heterogeneous treatment effects remains a wide-ranging debate across social sciences, medicine, economics, and other fields \citep{haushofer2022targeting,athey2025machine}. Choosing one or the other should involve an explicit discussion of trade-offs, values, and priorities --- rather than a choice made implicitly by adopting previously developed tools. 

{Organizations should also look beyond targeting, whether predictive or interventional, to improve outcomes by considering universal allocation, increasing resources, or intervention design. }
Ultimately, targeting is just one lever among many other available levers to improve outcomes. Differences between interventional and predictive targeting also warn that the considered intervention set may not be effective for the worst-off, calling for changes in intervention design to incorporate higher-touch interventions. 
Recent work illustrates how when it comes to improving social welfare, the bottom line impacts of improvements in predictive accuracy are outweighed by those of other policy levers such as expanding access \citep{perdomo2023relative,fischer2025value,fischer2026metadesign}. 

{Finally, besides choosing different decision-theoretic criteria, organizations may need to refine standard predictive problem formulations in view of statistical challenges from historical or future decisions.}
An organization's own prior decision-making processes may have shaped their historical data, and may require changes in predictive problem formulations. 
Collecting ground-truth outcome data can be difficult \citep{guerdan2023ground,zanger-tishler-risk-scores}. Sometimes historical decisions \textit{censor} the outcome observations, called ``selective labels"~\citep{lakkaraju2017selective}. For example, detaining defendant censors observing whether or not the defendant would have failed to appear or not if released. 
Beyond missingness, prior decisions may also causally affect observed outcomes. Since organizations adopt beneficial interventions that improve outcomes, naive ``zombie predictions" from the past may overestimate current risk \citep{koepke2018danger,hardt2023your}.
By accounting for prior historical interventions \citet{coston2020counterfactual,keogh2024prediction}, organizations can predict expected outcomes under each intervention, $\E[Y(1)\mid X], \E[Y(0)\mid X],$ which is more robust to future shifts in policy regimes. Finally, prior decisions imply the potential relevance of future decisions in social context. Recent research develops problem formulations beyond simple prediction that can account for future performativity~\citep{perdomo2020performative,perdomo2025making,kim2022making}
or users' strategic behavior~\citep{hardt2016strategic}.

\subsection{Evaluation science}

Evaluation should ask whether ADS change decisions and outcomes, not only whether they predict accurately. Our proposal entails that ADS evaluation needs to move beyond predictive benchmarking in light of historical and future decisions: the adoption of an integrated policy intervention framework enables the use of a wide menu of causal inference tools from quasi-experimental methods to randomized controlled trials to extend and innovate on existing ADS evaluation practice. 



{It should become standard practice to evaluate the causal impacts of ADS on downstream decisions and outcomes, and not just predictive error.}
ADS evaluation should include decision-centric evaluation criteria defined on the full integrated view $(R,\hat Y, D, Y)$, beyond predictive accuracy alone which only compares $(R, \hat Y, Y)$. Predictive accuracy alone rarely translates to improvement on the downstream outcomes of substantive interest which are otherwise omitted from standard reporting on predictive performance. 
Decision-centric evaluations can range from simple averages of downstream decisions and outcomes, to more difficult-to-estimate causal contrasts in outcomes from the same individual under different decisions.
For example, monitoring simple group-wise averages of who receives intervention can still give new policy insights: 
\citet{zhou2023optimal} studies supervised release and shows that alternate interventional targeting can improve racial disparities in electronic monitoring at relatively low cost: the room for improvement comes from changing the targeting strategy from predictive to interventional. Assessing whether those who strictly improve under decision receive intervention or not is more challenging \citep{kallus2019assessing, imai2023principal}. Yet, key moral facts and intuitions can be phrased as difficult counterfactuals. For example, understanding whether cash bail indeed provides specific deterrence in pretrial detention can be phrased as understanding whether someone who received bail and was then detained and unable to fail to appear, would have indeed failed to appear if they did not receive bail. 

Current impact evaluation methods do not reflect the full richness of our integrated view, and therefore cannot inform finer-grained decisions about ADS predictive model design vs. implementation science of ADS workflow design vs. intervention design. Some evaluations focus on the impact of $Z\to Y$ alone, i.e. outcomes before or after deployment of ADS, but therefore can only inform coarse go/no-go decisions for \textit{this} specific ADS. Regression discontinuity designs assess local causal impacts of high- or low-risk labels from predictive models at the threshold, but are therefore restricted to impact at the threshold alone. 
The framework of performative prediction emphasizes high-level dynamic modeling of how predictions shape the outcomes they are intended to predict \citep{perdomo2020performative,pastandfuture}, e.g. $\hat Y \to Y$, but without explicit articulation of underlying the treatment effect mechanisms as to how this happens. 
Current experimental and observational studies evaluating ADS tend to also 
under-consider how ADS model properties impact individual human \emph{decision responses} $\hat Y \to D$
\citep{raji2025evaluating}. For example, a high ADS positive prediction rate might lead to alert fatigue and lower human decision-maker responsiveness, especially in the context of capacity constraints. Similarly, a noticeably inaccurate ADS model can lower human trust, which could then cause decision-makers to more readily dismiss the ADS recommendations or ignore them altogether in their decision-making (i.e. minimizing or eliminating the observed causal effect). 

ADS differs from standard interventions in important ways that in turn should change evaluation design. First, we argue that ADS cannot themselves impact outcomes except by influencing decisions. Telling a decision-maker that the predicted risk of a future incident is 80\% cannot causally improve outcomes. ADS acts in the world by influencing decisions and interventions taken. Therefore, the impacts of ADS on outcomes should be disaggregated in terms of their impacts on decisions taken, and the impacts of different decisions taken on outcomes. For whom do ADS change decisions, and why? The answer can inform vastly different routes to improve outcomes: investigating the implementation science of human-ADS interaction and workflows ($R,D$) vs. revisiting intervention design ($D,Y$).
Second, causal impact evaluation of ADS is difficult because the ADS $R,$ is usually a non-randomized function of covariates, and therefore violates the \textit{overlap} (or positivity) assumption in causal inference. 
The \textit{policy change} in our integrated view introduces variation in who receives ADS that enables causal effect estimation.
For example, distinct pre- and post-implementation time periods enable event-study methods that compare outcomes before and after deployment \citep{albright2019if}. The ideal may be case-level experimental randomization across individual decision instances, as in \cite{imai2023experimental}, or across multiple decision-makers \citep{raji2025evaluating}. In the absence of ideal randomization, different quasi-experimental, experimental, or observational causal methods span different compromises between inferential validity and scope of relevance. Even then, tailoring methods to the specific nature of ADS can improve evaluation design. For example, disaggregating ADS impacts on decision take-up vs. impacts of decisions on outcomes \citep{zhou2023optimal} enables more informative robust policy optimization formulations under overlap violations \citet{ben2021safe,zhang2022safe}. 
Finally, thorough ADS evaluations may require organizations to modify data collection or merge different sources to track prediction, decision, and downstream outcome. Our integrated framework reveals important current information gaps in ADS evaluation. 
For example, New York City releases extensive information on its risk assessment tool's evaluation, but not how release decisions $D=0$ change as a result of variation in the predictive risk score or recommendation \citep{nycja-2025validation}. \textit{Separately,} pretrial data from ``judge IV" studies assess the \textit{causal impact of the RAT's deployment} on downstream outcomes $D \to Y$ \citep{koppel2024examining}, though not including any RAT-specific information. 

\subsection{Implementation science}


Finally, ~\emph{implementation science} focuses on the particularities of the deployment context --- i.e., determining how to best integrate and adopt ADS in practice. It is in the midst of layers of context  --- ranging from the individual to organizational to broader societal environment --- that the prediction and intervention actually interact to shape outcomes. Often we confine our modeling of contextual factors to the hyper-local considerations of individual user decision-making, however a more grounded perspective involves a broader view, incorporating information about the institutional, societal and legal environment into which the ADS system is released into our understanding of how an ADS deployment may impact downstream outcomes.

For example, consider the prediction tasks of (A) criminal recidivism and (B) hospital re-admission. Although near identical in problem formulation (i.e. how likely is the release/discharge to fail and the model subject to return? \footnote{Note that there exist similar tasks in other domains such as child-welfare alert algorithms~\citep{capatosto2017foretelling}, and homeless service prioritization algorithms such as VI-SPDAT~\citep{petry2021associations}}), data inputs and modeling strategies, these predictions qualitatively mean very different things to the organizations that choose to adopt them, and the decision-makers implementing interventions in-situ~\citep{saxena2023}, in large part due to various individual, organizational and normative societal contextual factors. 

If implementation factors are considered at all, computer scientists are used to thinking about this in a hyper-local \emph{individual} user context level, where ~\emph{user interaction design} (i.e. how $R$ and $\hat{Y}$ are presented to, and interpreted by the individual decision-maker to inform decisions $D$) and ~\emph{user intervention design} (i.e. the set of available actions  $D$ an individual decision-maker can take in response to a particular prediction or assessment) are the primary axes of consideration. 

This is a warranted concern -- the interaction between an ADS and its human decision-maker should be treated as a central design choice, not a side effect. Different deployments require different balances of control and information: some settings call for full automation when the model and human have access to the same information and errors are costly; others require decision support, where predictions are interpretable and uncertainty is shown so that humans can rely on them appropriately. Display, timing, and presentation format of the prediction impact decision take-up. When models and humans hold different information, systems should learn to defer---routing routine cases to the algorithm and ambiguous ones to people---or, when human oversight must remain, communicate comparative reliability so the person knows when to trust the model. The key design variable is therefore not whether humans are ``in the loop,” but how the loop is structured to achieve complementary, calibrated decisions.

Yet, in most social settings, individual users rarely operate with full human discretion. Rather, they operate within a larger \emph{organizational} context, governed by its own constraints and additional considerations. These factors at the organizational level can completely shape the impact of prediction outcomes. 
For example, multiple stakeholders have legitimate claims to shape deployment scenarios, but each stakeholder may have different incentives and priorities which in turn may not align with the decision subject \citep{laufer2023strategic}. Capacity constraints and resource limitations can impact everything from user responsiveness to intervention effectiveness~\citep{klein2017sources, boutilier2024randomized, raji2025evaluating, buccinca2021trust, goddard2012automation}. 

Furthermore, various implementation factors 
are likely to be determined at an industry or ~\emph{societal} scale. 
Because the underlying bureaucratic counterfactual policy for decision-making is typically under the purview of some judicial actor, any prediction-mediated decision-making will likely be considered interchangeable with that historical status quo, and thus fall under the same legal and regulatory jurisdiction. 
Therefore, prediction-based decision-making can be subject to similar legal scrutiny. 
In fact, the legal compatibility of evaluation and implementation is necessary for a predictor to be operable in deployment, and for the decisions made with such tools to be deemed legally valid~\citep{xiang2019legal, raghavan2024limitations}. 

Currently, the field tends to underestimate the complexity of the layers of individual, organizational and societal context in which ADS systems are integrated and deployed. Most deployments manifest as multi-prediction, multi-stakeholder pipelines --- multiple models used at different stages of a complex decision-making process, interacting with distinct stakeholders and other models as part of a much broader ecosystem. 
Although some deployment studies do explicitly attempt to document the process of organizational workflow integration \citep{callahan2024standing, sendak2020real,saxena2021admaps, kim2023organizational,callahan2024standing}, mapping such operational dynamics is far from the norm.

\section{Conclusion}


Prediction is just one of 
many paths to improving downstream outcomes. 
When predictions are deployed, they themselves act as interventions by affecting decisions which affect outcomes, as well as procedural fidelity of consequential decision processes. 
The first step towards making normative conclusions about whether ADS-informed decisions improve social system outcomes is to measure their full effects within a specific context or use, including for their procedural effects. 
We have highlighted how these composite ways that ADS acts in the world should change how developers design, evaluate and implement ADS by revisiting
fundamental sociotechnical constraints, 
success criteria, 
and new design affordances. 

\bibliographystyle{abbrvnat}

\bibliography{draft4-forarxiv-clean}

\end{document}